\documentclass[aps,prb,preprint,superscriptaddress,amsmath,amssymb]{revtex4-2}
\usepackage{CJKutf8}
\usepackage{graphicx}
\usepackage{siunitx}
\usepackage{amsmath}

\def\bea{\begin{eqnarray}}
	\def\eea{\end{eqnarray}}
\def\ben{\begin{equation}}
	\def\een{\end{equation}}
\def\benu{\begin{enumerate}}
	\def\enu{\end{enumerate}}
\def\beal{\begin{equation}\begin{aligned}} 
		\def\eeal{\end{aligned}\end{equation}}

\def\sss{\scriptscriptstyle\rm}

\def\hatH{{\hat H}}

\def\br{{\mathbf r}}
\def\brp{{\mathbf r}^\prime}
\def\tp{t^\prime}

\def\bq{{\mathbf q}}

\def\bj{{\mathbf j}}

\def\cF{\mathcal{F}}

\def\diff{{\rm d}}
\def\d3{{\rm d^3}}

\def\intdbr{\int\diff{\br}}

\def\s{_{\sss S}}
\def\b{_{\sss B}}

\def\P{_{\sss P}}

\def\VW{_{\sss vW}}

\def\nad{^{\rm nad}}
\def\ad{^{\rm ad}}

\usepackage{xspace}

\newcommand{\etal}{{\it et al.}\xspace}

\usepackage{mathtools}

\newcommand{\eqn}[1]{\mbox{Eq.\hspace{1pt}(\ref{#1})}}

\begin{document}
	
\begin{CJK*}{UTF8}{gbsn}
\title{Nonlocal and nonadiabatic Pauli potential for time-dependent orbital-free density functional theory}

\author{Kaili Jiang (姜凯立)}
\email{kaili.jiang@rutgers.edu}
\affiliation{Department of Chemistry, 73 Warren St., Rutgers University, Newark, NJ 07102, USA}

\author{Xuecheng Shao (邵学成)}
\email{xs161@newark.rutgers.edu}
\affiliation{Department of Chemistry, 73 Warren St., Rutgers University, Newark, NJ 07102, USA}

\author{Michele Pavanello}
\email{m.pavanello@rutgers.edu}
\affiliation{Department of Chemistry, 73 Warren St., Rutgers University, Newark, NJ 07102, USA}
\affiliation{Department of Physics, 101 Warren St., Rutgers University, Newark, NJ 07102, USA}

\begin{abstract}
	Time-dependent orbital-free density functional theory (TD-OFDFT) is an efficient ab-initio method for calculating the electronic dynamics of large systems. In comparison to standard TD-DFT, it computes only a single electronic state regardless of system size, but it requires an additional time-dependent Pauli potential term. We propose a nonadiabatic and nonlocal Pauli potential whose main ingredients are the time-dependent particle and current densities. Our calculations of the optical spectra of metallic and semiconductor clusters indicate that nonlocal and nonadiabatic TD-OFDFT performs accurately for metallic systems and semiquantitatively for semiconductors. This work opens the door to wide applicability of TD-OFDFT for nonequilibrium electron and electron-nuclear dynamics of materials.
\end{abstract}

\date{\today}
\maketitle
\end{CJK*}

\section{Introduction}
The ab-initio simulation of nano-scale quantum systems when their electrons are out of equilibrium is a challenging task \cite{SD07,HCM08,MLB08,SD08,Vignale08}, not only for time-dependent DFT (TD-DFT) but for almost all other electronic structure methods. One of the challenges is the computational cost involved in the simulations, which compounds on top of the already challenging computation of the ground state. In DFT and TD-DFT, the computational expense typically scales cubically with system size, and for wavefunction-based methods, the cost is even higher \cite{PJ12}.  The ab-initio methods currently available, are thus not suited to address systems relevant to most experiments (where typical sizes of nanoclusters are in the range of 1 to 1000 nm \cite{Ringe20,Baletto19,AGW+16}).  Clearly,  alternative methods are needed. Several research groups are actively working to tackle this issue developing such methods as the semi-empirical time-dependent density functional tight biding (TD-DFTB)\cite{RLH+16}. In this work, we further the time-dependent orbital-free density functional theory (TD-OFDFT), an {\it ab initio} method with potentially far reduced computational cost compared to TD-DFT and even TD-DFTB.

TD-OFDFT tackles the scaling problem by only considering one active orbital. In contrast, TD-DFT uses a number of orbitals equal  to the number of electrons in the system. Thus, TD-OFDFT can achieve a linear computational scaling with the number of electrons so long as approximate functionals are used. In addition to the time-dependent exchange-correlation (XC) potential that also needs to be approximated in TD-DFT \cite{FLN+18}, TD-OFDFT requires to approximate the time-dependent Pauli potential \cite{JP21,WCD+18,DWH+18,SG17}. Thus, the main challenge is to find a good approximation to this potential.

Early adoptions of TD-OFDFT utilize adiabatic and local density approximations, i.e., the adiabatic Thomas-Fermi (TF) \cite{Thomas26,Fermi27} potential as the time-dependent Pauli potential, sometimes referred to as time-dependent Thomas Fermi  \cite{PBN+86,DRS98,RLB00}, or hydrodynamic DFT \cite{BH00,MGB15,MBR18,Yan15}. Such a simple approximation has been  successful for systems such as Na clusters \cite{DRS98}, but only qualitatively captures the spectra for other types of systems \cite{BH00,PB15,SJM+21}. As we will show in this work, TD-OFDFT can only become semiquantitative for a wide class of systems if the approximate time-dependent Pauli potential adopted in the simulations is nonlocal and nonadiabatic.

To account for nonlocality, several nonlocal kinetic energy density functionals (KEDFs) have been developed for ground state OFDFT \cite{HC10,MGP18,MP19}. These can be employed in TD-OFDFT by applying the so-called adiabatic approximation (i.e., the potentials are given by the ground state functional evaluated at the time-dependent electron density). To account for nonadiabaticity (i.e., going beyond the adiabatic approximation),  potentials need to become at least dependent on the time-dependent electron density as well as the current density \cite{VK96,VUC97,FLN+18}, as done recently by White \etal\  \cite{WCD+18}. Along these lines of research, in this work we derive an improved  nonadiabatic correction to the time-dependent Pauli potential from a frequency-dependent Pauli kernel derived from the response of the uniform electron gas (UEG) which we recently proposed \cite{JP21}.

Another challenge for TD-OFDFT has been the general lack of software that implements it. We recently filled this gap by developing DFTpy \cite{SJM+21} (dftpy.rutgers.edu), an object-oriented Python software for OFDFT as well as real-time and linear-response (Casida \cite{Casida95}) TD-OFDFT. DFTpy is fully parallelized using mpi4py \cite{DPS+08}, and scales well with system size as we showed in Ref.\ \citenum{SJM+21}.

This work is structured as follows. In Section \ref{sec:formalism} we briefly introduce the formalism of TD-OFDFT, derive the proposed nonadiabatic correction to the time-dependent Pauli potential from the frequency-dependent Pauli kernel, and discuss the implementation of the potential in DFTpy. In Section \ref{sec:results} we present the optical spectra derived from real-time TD-OFDFT runs carried out in a linear-response regime (weak perturbation). As examples, we chose several metallic (Na, Mg, Ag) and semiconductor clusters (Si, GaAs). %We show that by including  nonlocality and nonadiabaticity in the time-dependent Pauli potential, TD-OFDFT achieves semiquantitative agreement with the spectra from full TD-DFT simulations.

%hard to compute many states, have to smear many states, people looking at TD-DFTB, non-ab-initio.
%TD-OFDFT (we don't have a code before, but we have one now)
% Looking at different v_p

%In this work we are going to investigate the following systems: Na clusters, in which the TD-OFDFT results closely matches that of TD-DFT and the experiments; Ag nanorods, a system that ground-state OF-DFT doesn't perform particularly well, but TD-OFDFT is on par with semi-empirical method TD-DFTB, and close to the results of TD-DFT; Si and GaAs nanoclusters, in both cases TD-OFDFT yields qualitatively accurate results comparing with TD-DFT, gives the same position of the peaks and the same effect from passivation.

\section{Formalism}\label{sec:formalism}

\subsection{Time-dependent Schr\"{o}dinger-like equation for TD-OFDFT}

For more details on TD-OFDFT, we refer the readers to Refs.\ \citenum{JP21,WCD+18,Schaich93}. Here we provide a brief introduction to the formalism.
In TD-OFDFT, one-to-one invertible maps are established between the real system of interacting electrons, the fictitious system of noninteracting electrons (a.k.a.\ the KS system), and a fictitious system of noninteracting bosons. The bosonic system yields the following density-wavefunction relationship
\begin{equation}
	\label{}
	n(\br,t)=N|\phi\b(\br,t)|^2,
\end{equation}
where the subscription $\mathrm{B}$ stands for the bosonic system.

In a typical setup for real-time propagations, the system starts with its ground state density $n_0(\br)$ and at $t=0$ the system is perturbed with a kick with strength $k$ \cite{YNI+06}. A time-dependent Schr\"{o}dinger-like equation is employed to propagate the system for $t>t_0$, 
\begin{equation}
\label{eq:schr}
\left[-\frac{\nabla^2}{2}+v\b[n](\br, t)\right]\phi\b({\br, t})=i\frac{\partial}{\partial t}\phi\b({\br, t}),
\end{equation}
with the initial condition
\begin{equation}
\phi\b(\br,t_0)=\frac{1}{\sqrt{N}}\sqrt{n_0(\br)}e^{ikz},
\end{equation}
where the exponential term effectively donates momentum $k$ to the electronic system.
The time-dependent effective potential is given by
\begin{equation}
v\b[n](\br, t) = v\P[n](\br,t) + v\s[n](\br,t),
\end{equation}
where $v\P[n](\br,t)$ is the time-dependent Pauli potential which is the difference between the bosonic noninteracting kinetic potential and the fermionic one. Thus, it is said that the Pauli potential introduces the effects of the  Pauli exclusion prnciple in the boson electronic structure \eqn{eq:schr}. In practice, the Pauli energy is defined as the subtraction of the von Weizs\"{a}cker\cite{Weizsaecker35} (vW, $T\VW[n]$) functional from the exact noninteracting kinetic energy functional. Therefore, kinetic energy functionals developed for ground-state OFDFT can be used for approximating the Pauli potential by simply taking the functional derivative of the functional removed of the von Weizs\"acker kinetic energy. 

The corresponding adiabatic approximation for the Pauli potential is given by,
\begin{equation}
	\label{}
v\P\ad(\br,t)=\left.\frac{\delta T\P[n_0]}{\delta n_0}\right|_{n_0(\br)\rightarrow n(\br,t)},
\end{equation}
where $T\P[n_0]$ is the Pauli energy, $T\s[n_0]-T\VW[n_0]$. However, to achieve accurate results in TD-OFDFT, the nonadiabatic contribution to the potential should not be neglected. In this case, the total Pauli potential can be represented as the adiabatic portion plus a nonadiabatic correction
\begin{equation}
	\label{}
	v\P(\br,t)=v\P\ad(\br,t)+v\P\nad(\br,t).
\end{equation}

The nonadiabatic contribution to the Pauli potential has much stronger effects on the electron dynamics compared to the nonadiabatic contribution to the exchange-correlation (XC) potential \cite{FLN+18}. The latter, needs to bridge the dynamics of $N$ noninteracting electrons with the one of $N$ interacting electrons. Thus, states described by so-called double excitations and beyond need to be accounted for by the nonadiabatic XC term. In contrast, the nonadiabatic contribution to the Pauli potential needs to bridge the dynamics of $N$ noninteracting bosonic electrons to the one of $N$ noninteracting fermionic electrons, therefore it needs to create all those excitations that are missing when exciting a single effective electron in comparison to exciting $N$ fermionic electrons. Thus, the nonadiabaticity in the Pauli potential is very strong and typically accounts for qualitative aspects of the electronic response, as we reported recently \cite{JP21} in a study about the first excited state of selected systems. Thus, we expect the addition of an approximate nonadiabatic Pauli potential contribution to the TD-OFDFT effective time-dependent potential will significantly improve the description of the electronic structure of the systems considered in comparison to adiabatic TD-OFDFT.

\subsection{Approximating the nonadiabatic Pauli potential}

In this section, we will derive a nonadiabatic correction to the Pauli potential from a nonadiabatic Pauli kernel developed by the authors \cite{JP21}.  In our previous work, we derived a Pauli kernel relying on the Dyson equation connecting the bosonic response, $\chi\b$, with the Kohn-Sham response, $\chi\s$,
\begin{equation}
	\label{eq:bks}
	\chi\b^{-1}-\chi\s^{-1}=f\P.
\end{equation}
All the above quantities have a dependence on time as $t-\tp$ and two space variables, $\br$ and $\brp$. By substituting in \eqn{eq:bks} the frequency dependent Lindhard function for $\chi\s$ and the response of a uniform boson gas for $\chi\b$, one obtains an expression for $f\P$ which can be expanded in terms of unitless variables, $\bar\eta=\frac{q}{2k_F}$, $\bar\omega=\frac{\omega}{qk_F}$, $\bar\gamma=\frac{\eta}{qk_F}$, where $q$ is the conjugate variable to $|\br-\brp|$, $\omega$ is the frequency of the external time-dependent perturbation and $\eta$ is the broadening parameter introduced to implement causality \cite{Ullrich11}.

Retaining terms up to the first order of $\omega$,
\begin{equation}
	\label{eq:fp_nad}
	f\P\nad(\br,q,\omega)=\frac{i\pi^3}{12}\left(\frac{6}{k_F^2(\br)q}+\frac{q}{k_F^4(\br)}\right)\omega,
\end{equation}
where the local density approximation for the Fermi wavevector is introduced, $k_F(\br)=(3\pi^2n(\br))^{1/3}$.

We now use the nonadiabatic Pauli kernel to approximate the nonadiabatic correction to the time-dependent Pauli potential adopting the same technique as described in Ref.\ \citenum{WCD+18}.
Considering the continuity equation
\begin{equation}
	\label{eq:delta_n}
	\frac{\partial n(\br',t)}{\partial t} = -\nabla\cdot\bj(\br',t),
\end{equation}
and expanding the density at $t'$ around time $t$ up to the first order, we obtain
\begin{align}
	\label{}
	\delta n(\br',t')&\approx \delta n(\br', t) + \frac{\partial n(\br', t)}{\partial t}(t-t')\nonumber\\
	&= \delta n(\br', t) - \nabla\cdot\bj(\br',t)(t-t').
\end{align}
Using the definition of the nonadiabatic Pauli kernel
\begin{equation}
	\label{}
	f\P\nad(\br, \br', t-t') = \frac{\delta v\P\nad(\br,t)}{\delta n(\br',t')},
\end{equation}
we can write the variation of the Pauli potential as
\begin{equation}
	\label{eq:delta_vp}
	\delta v\P\nad(\br, t) = \intdbr'\,\int\diff(t-t') \delta n(\br',t')f\P\nad(\br, \br', t-t').
\end{equation}
Plugging \eqn{eq:delta_n} into \eqn{eq:delta_vp} we get
\begin{align}
	\label{eq:vp_nad}
	v\P\nad(\br, t) &= -\intdbr'\,\nabla\cdot\bj(\br',t)\int\diff(t-t') (t-t')f\P\nad(\br, \br', t-t')\nonumber\\
	&=-\intdbr'\,\nabla\cdot\bj(\br',t)\frac{i\partial f\P\nad(\br,\br',\omega)}{\partial \omega}\nonumber\\
	&= -\cF^{-1}\left\{i\bq\cdot\bj(\bq,t)\frac{i\partial f\P\nad(\bq,\omega)}{\partial \omega}\right\},
\end{align}
where $\cF^{-1}\{\cdot\}$ represents inverse Fourier transform in the space domain. 

We immediately clarify that \eqn{eq:vp_nad} is approximate. Not only because the time-dependent density is expanded only to first order, but, most importantly, because the recovered quantity is formally not 
$v\P\nad$ 
but its first-order variation $\delta v\P\nad$. Formally, a procedure of functional integration should be employed which, unlike the static case \cite{LB95,GCS09,MGP18}, is not straightforward in the time domain.

Finally, we plug \eqn{eq:fp_nad} into \eqn{eq:vp_nad} and note that the derivative of the adiabatic contribution of $f\P$ with respect to $\omega$ vanishes, we obtain
\begin{equation}
	\label{eq:vp_jp}
	v\P\nad(\br,t) = \frac{\pi^3}{12}\left(\frac{6}{k_F^2(\br)}\cF^{-1}\left\{i\bq\cdot\bj(\bq,t)\frac{1}{q}\right\}+\frac{1}{k_F^4(\br)}\cF^{-1}\left\{i\bq\cdot\bj(\bq,t)q\right\}\right).
\end{equation}
We note that \eqn{eq:vp_jp} corrects the potential proposed by White \textit{et al} \cite{WCD+18} with the addition of a second term. As our calculations will demonstrate, the second term of \eqn{eq:vp_jp} is key to achieving accurate results and thus should not be neglected.

\subsection{Implementation of the nonadiabatic potential}

We numerically solve \eqn{eq:schr} in the usual way by a propagation method implemented in our software package DFTpy. We choose the Crank-Nicolson propagator for this task, with a predictor-corrector up to any desired order. We refer the readers to Ref.\ \citenum{SJM+21} for the detailes of the implementation. In short, the propagator is to solve the following equation
\begin{equation}
\label{eq:cn}
\left(1+i\frac{\diff t}{2}\hat{H}\right)\phi\b(t+\diff t)=\left(1-i\frac{\diff t}{2}\hat{H}\right)\phi\b(t).
\end{equation}

Typically, the predictor-corrector only checks the density to be converged \cite{CMR04,GMR+18}. However, to implement a current-dependent potential such as the one in \eqn{eq:vp_jp}, the predictor-corrector needs to be modified to include the current density, i.e., $|j_{\rm{corr}}-j_{\rm{pred}}|<\epsilon$. Unfortunately, this  often results in a large number of predictor-corrector loops, increasing computation time and causing numerical instabilities. To resolve this issue, we implemented the nonadiabatic potential as a correction with a separate Taylor-like propagation.

To achieve this goal, we first write the Hamiltonian as
\begin{equation}
\hat{H}(t) = \hat{H}^0(t) + \hat{H}'(t),
\end{equation}
where $\hat{H}'$ includes the nonadiabatic potential in \eqn{eq:vp_jp} and $\hat{H}^0$ includes everything else. The solution of the time-dependent Schr\"{o}dinger-like equation \eqn{eq:schr} can be expressed in terms of a time evolution operator \cite{GMR+18}
\begin{equation}
\label{eq:te_total}
\phi\b(t+\diff t)\approx\exp\left[-i\hatH(t+\frac{\diff t}{2}) \diff t\right]\phi\b(t),
\end{equation}
where the approximation implies a small $\diff t$.

In the first step, we use the Crank-Nicolson propagator with a predictor-corrector for only $\hat{H}^0(t)$. We essentially obtain
\begin{equation}
\label{eq:te_ad}
\phi\b^0(t+\diff t)\approx\exp\left[-i\hatH^0(t+\frac{\diff t}{2}) \diff t\right]\phi\b(t).
\end{equation}

Plug \eqn{eq:te_ad} into \eqn{eq:te_total}, we obtain
\begin{equation}
\label{eq:te_nad}
\phi\b(t+\diff t)\approx\exp\left[-i\hatH'(t+\frac{\diff t}{2}) \diff t\right]\phi\b^0(t+\diff t).
\end{equation}
Note here we used the approximation $\exp\left[-i\hatH(t+\frac{\diff t}{2}) \diff t\right]\approx\exp\left[-i\hatH^0(t+\frac{\diff t}{2}) \diff t\right]\exp\left[-i\hatH'(t+\frac{\diff t}{2}) \diff t\right]$. Even though $\hat{H}'$ does not commute with $\hat{H}^0$, the cross terms are in the second and higher order of $\diff t$. In the case of small $\diff t$, the cross terms are negligible.

In the case of small $\hat{H}'(t)$ and small $\diff t$, we can make the assumption that $\hatH'(t+\frac{\diff t}{2})\approx\hatH'(t)$ and Taylor expand \eqn{eq:te_nad} up to the first order
\begin{equation}
\phi\b(t+\diff t)=\left[1-i\hatH'(t)\diff t\right]\phi\b^0(t+\diff t).
\end{equation}
Therefore, as the second step, we use the first-order Taylor propagator to propagate the nonadiabatic correction. We tested this method for a Mg$_8$ cluster and achieved nearly identical result compared to propagating the whole Hamiltonian with the Crank-Nicolson propagator.

We note that when using \eqn{eq:vp_jp} for clusters, the $k_F^{-4}(\br)$ dependence of the second term can cause numerical instabilities in low density regions. In such case, a mask function such as $1-1/\{1+[n(\br)/n{\sss cutoff}]^2\}$ can be applied to the second term to fix the numerical instabilities.

\section{Computational Details}
The TD-OFDFT calculations are performed with DFTPy \cite{SJM+21} and the benchmark TD-DFT calculations are performed with embedded Quantum Espresso (eQE) \cite{GCK+17,MSG+21}  and its real-time TD-DFT implementation \cite{KCP15}. All calculations employ the same adiabatic Perdew-Zunger LDA \cite{PZ81} as the XC functional and OEPP \cite{MZW+16} pseudopotentials. The kinetic energy cutoff for the TD-DFT wavefunctions and the TD-OFDFT density are chosen to converge the ground state energy within 1 meV/atom. The oscillator strength is calculated using $\sigma(\omega)=-\omega\text{Im}\bigg[\delta\tilde{\mu}(\omega)\bigg]$, where $\delta\tilde{\mu}(\omega)$ is the Fourier transform of the dipole moment change calculated at every time step of the propagation. The time step chosen is consistent with the plane wave cutoff employed and is $\diff t=0.1$ a.u. in all simulations, except for Ag, Si and GaAs simulations with the LMGP functional, where the time step is $\diff t=0.01$ a.u.

\section{Results and Discussion}\label{sec:results}

We first present results for clusters of metallic systems (Mg, Na, Ag) and in a second step we discuss results for  clusters of semiconducting systems (Si, GaAs).

\subsection{Benchmarking metallic systems}
%\subsection{Mg}

\begin{figure}[htp]
	\centering
	\includegraphics[width=\textwidth]{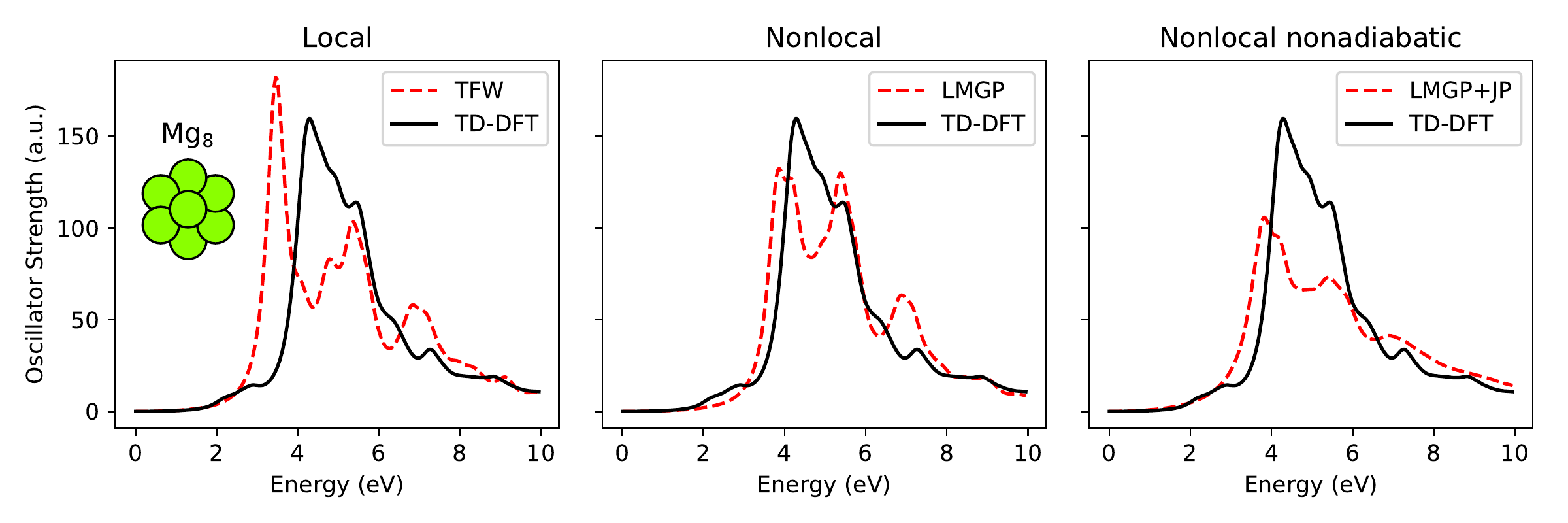}
	\caption{Optical spectrum of the Mg$_8$ cluster (see inset for a snapshot of the structure). LMGP is a nonlocal kinetic energy functional with density dependent kernel \cite{MP19} employed in the adiabatic approximation (i.e., $v\P(t)\simeq v\P[n_0=n(t)]$). LMGP+JP stands for adiabatic LMGP augmented by the additional propagation step with the nonadiabatic potential in \eqn{eq:vp_jp}.}
	\label{fig:Mg8}
\end{figure}

Previously \cite{SJM+21}, we investigated the TD-OFDFT optical spectra of Mg clusters with the adiabatic Thomas-Fermi approximation for the Pauli potential (TFW here onward). Our results showed that TFW is only in semiqualitative agreement with the TD-DFT benchmark. We remark that we take TD-DFT as benchmark here, because it formally has an exact Pauli potential. Figure \ref{fig:Mg8} shows that TD-OFDFT can reach quantitative agreement with TD-DFT provided nonlocality (with the LMGP kinetic energy functional) and nonadiabaticity (adding to LMGP the nonadiabatic contributions to the Pauli potential) are included.

The effect on the optical spectrum of Mg$_8$ of including nonlocality in the Pauli potential is very strong. TFW significanlty red-shifts the first absorption band compared to the benchmark TD-DFT results. Accounting for nonlocality in the Pauli potential by employing the adiabatic LMGP \cite{MP19} nonlocal potential brings the spectrum closer to the TD-DFT benchmark. However, the spectral envelope is not recovered fully. When the nonadiabatic correction in \eqn{eq:vp_nad} (JP here onward) is included, then TD-OFDFT recovers the TD-DFT spectral envelope despite slightly increasing the spectral weights in the tail of the spectrum at high energies depleting of intensity the main spectralfeatures. 

%It also shows that the CD potential corrects the relative strength of the peaks in the opposite direction, resulting in a worse spectra. This indicates the importance in the nonadiabaticity of the Pauli potential for large $q$ which is taken account for by the second term of \eqn{eq:vp_nad}.

\begin{figure}[htp]
	\centering
	\includegraphics[width=\textwidth]{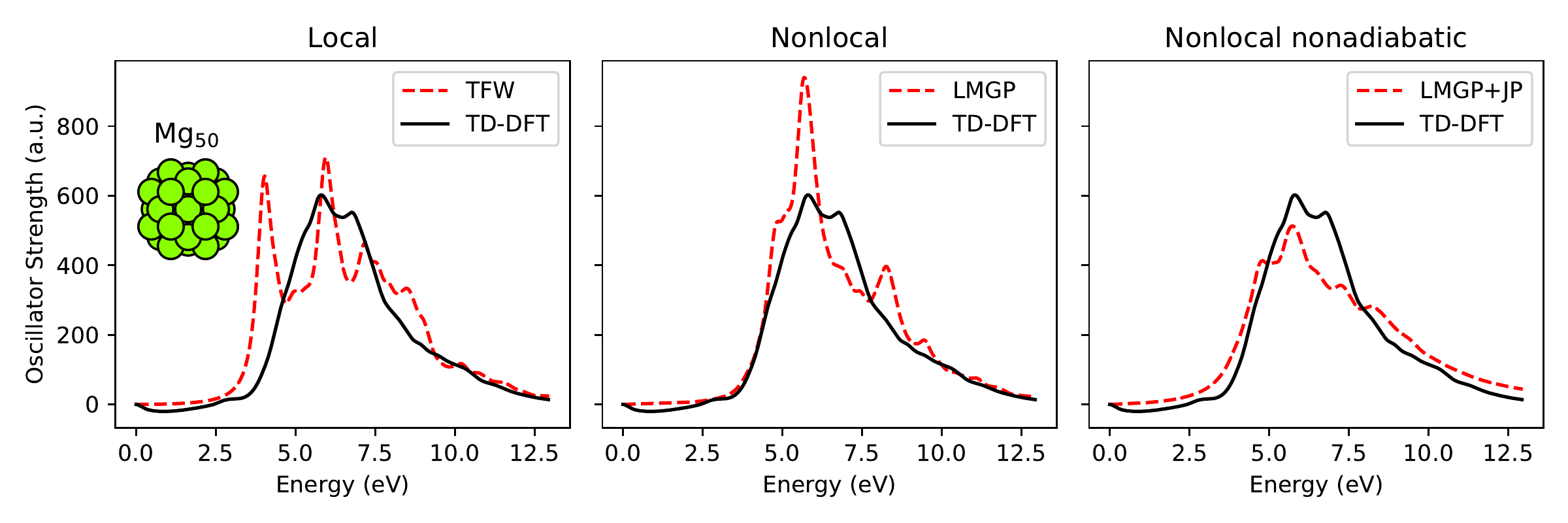}
	\caption{Optical spectrum of the Mg$_{50}$ cluster. See caption to Figure \ref{fig:Mg8} for additional details.}
	\label{fig:Mg50}
\end{figure}

In Figure \ref{fig:Mg50} we report the computed optical spectrum of a Mg$_{50}$ cluster. The inclusion of the nonlocality and the nonadiabaticity in the Pauli potential has very similar effects in the spectra to that of the Mg$_8$ cluster, with the nonlocal term shifting the peaks to the correct position and the nonadiabatic term adjusting the shape of the spectral envelope. It is also worth noting that the TD-OFDFT results get more accurate as the system grows larger, as the larger system is more likely to develop free-eletron-gas like electronic structure in its core \cite{SJM+21}.

%\subsection{Na}

\begin{figure}[htp]
	\centering
	\includegraphics[width=\textwidth]{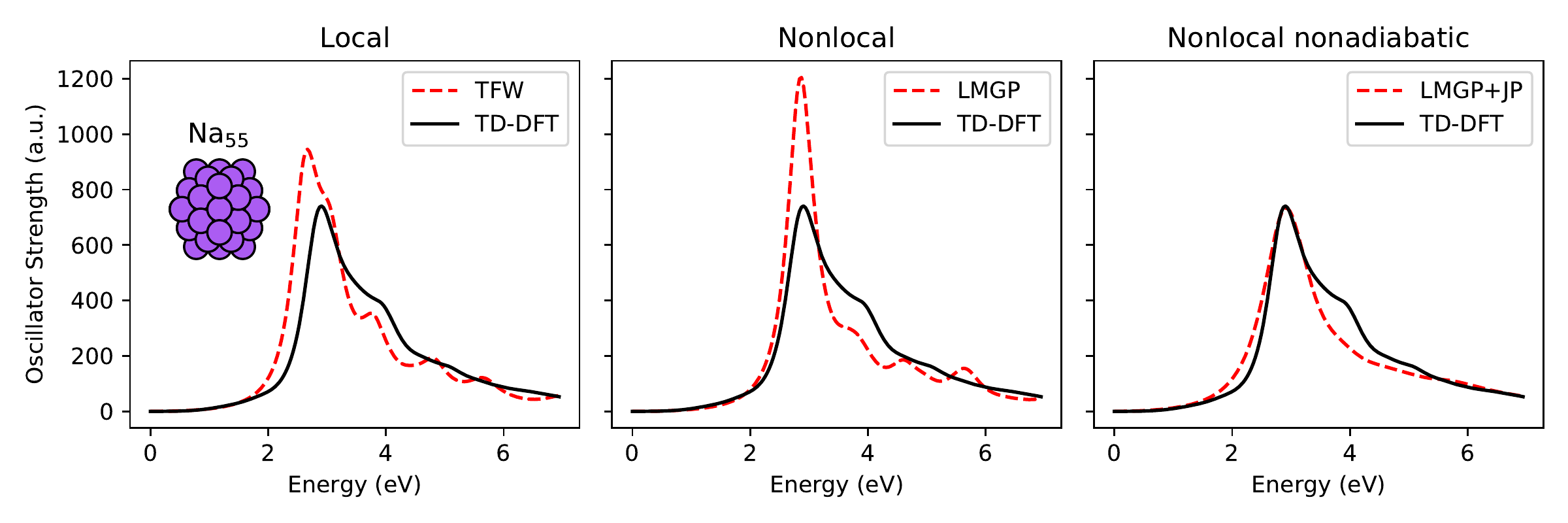}
	%\caption{Comparison of the effects of the nonlocal and nonadiabatic Pauli potential in the oscillator strength for icosahedron Na$_{55}$ cluster. The insect shows the structure of the cluster. Black solid: TD-DFT benchmark. Red dash-dotted: TFW (local). Green dashed: LMGP (nonlocal). Blue solid: LMGP+JP (nonlocal and nonadiabatic).}
	\caption{Optical spectrum of the Na$_{55}$ cluster. See caption to Figure \ref{fig:Mg8} for additional details.}
	\label{fig:Na55}
\end{figure}

We now shift to Na clusters. In all past investigations even the simplest of approximations for the Pauli potential (i.e., TFW) resulted in oscillator strengths that  agree well with the TD-DFT result \cite{DRS98}. In light of the simulations presented here, it is clear that such agreement is largely due to error cancellation between the missing adiabatic nonlocal term and the missing nonadiabatic correction for the Pauli potential in TFW. In this work, we take a look at a icosahedron Na cluster containing 55 Na atoms, see Figure \ref{fig:Na55}.

%Fig. \ref{fig:Na55} compares the oscillator strength between the TD-OFDFT real-time, TD-OFDFT Casida and TD-DFT for the icosahedron Na$_{55}$ cluster. The shape and the position of the peak is almost identical for both methods and the intensity of the peaks are similar. This great agreement between TD-OFDFT and TD-DFT in Na, is mostly due to the fact that each Na atom contains a single valence electron. Thus the eletronic structure of the Na cluster can be accurately represented by the non-interacting Boson model even with a simple adiabatic TF Pauli potential. In terms of the comparison between the two TD-OFDFT methods, the results are mostly identical and the slightly different intensity is due to the different smearing method used.  

As hinted, the spectrum calculated with TFW already matches well with the TD-DFT benchmark with only a slight red shift, confirming previous results \cite{DRS98}. When accounting for the nonlocal correction to the adiabatic Pauli potential computed at the LMGP level,  peaks shifts to the correct position but the strength of the main peak at $\sim 5$ eV is overestimated. Only after including both nonlocal and nonadiabatic corrections, does  TD-OFDFT yield a spectral envelope that is very close the one from TD-DFT. Thus, it is clear that error cancellation is a big player in the previously reported good results from the TFW approximation for the Pauli potential.

%\begin{figure}[htp]
%	\centering
%	\includegraphics[width=0.70\textwidth]{figure/Na.pdf}
%	\caption{The oscillator strength for icosahedron Na clusters with different sizes. The insect shows the structure of the Na$_{2057}$ cluster.}
%	\label{fig:Na}
%\end{figure}

%Fig. \ref{fig:Na} compares the oscillator strength for 3 different sized icosaherdron Na clusters with TD-OFDFT. The latter two clusters, Na$_{309}$ and Na$_{2057}$, are too large for TD-DFT to complete in a reasonable amount of time but can be done with ease for TD-OFDFT thanks to its linear scaling. The result shows the different-sized icosaherdron Na clusters yields very similar oscillator strength per electron.

%\subsection{Ag}

\begin{figure}[htp]
	\centering
	\includegraphics[width=0.584\textwidth]{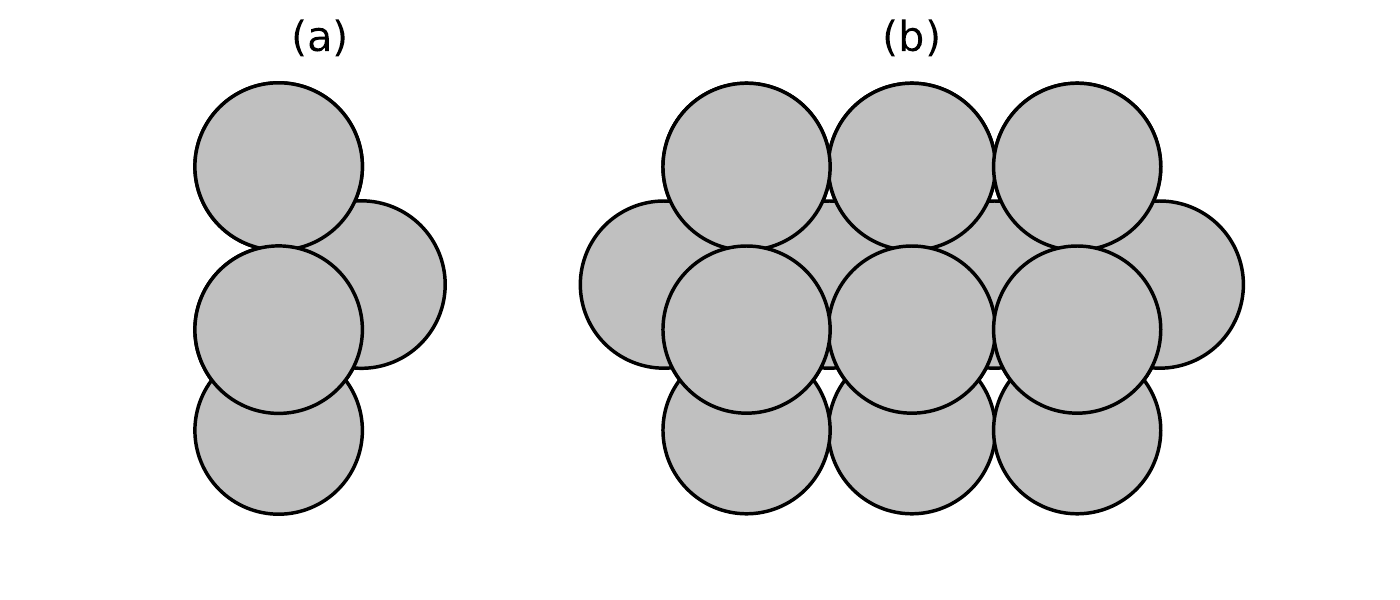}
	\caption{(a) Ag$_n$ nanorod building block. (b) $n=19$ nanorod. The procedure for building the Ag nanorods is taken from Refs.\ \citenum{JA09, AA18}.}
	\label{fig:Ag}
\end{figure}

Silver is an important transition metal with high applicability to chemistry and engineering. Thus, its electronic response has often been the subject of approximate models\cite{MZW+16,AA18}. We compute the longitudinal plasmon excitation of Ag nanorods and compare against TD-DFT and the semi-empirical time-dependent density functional tight binding (TD-DFTB) presented in Ref.\ \citenum{AA18}. Snapshots of the Ag nanorods used can be found in Figure \ref{fig:Ag}. 

%Fig. \ref{fig:Ag_rod} Compares the oscillator strength of the Ag$_{55}$ nanorod among the three methods. 
Even though the TD-DFT and TD-DFTB calculations were carried out with the Perdew-Burke-Ernzerhof (PBE) XC functional and the DZ basis set \cite{AA18} which differ from the setup of our TD-OFDFT calculations (we employ LDA XC and plane waves), the trends in the computed excitation energies should not be strongly dependent on the choice of local or semilocal XC functionals  \cite{WXK12}. Table \ref{tab:Ag_rod} shows that TD-OFDFT with TFW is very close to the TD-DFTB result and LMGP+JP  is essentially on top of TD-DFTB. In inspecting the table, we see that TD-OFDFT correctly captures the trend that the peaks shift to the red as the Ag rod's length increases. 

\begin{table}[h]
	\caption{Comparison of the excitation energy of the longitudinal plasmon peaks of Ag rods of increasing length. TD-DFT and TD-DFTB are taken from Ref.\ \citenum{AA18}. The short hand notation ``pos'' and ``shift'' stand for peak position (excitation energy) and shift from the Ag$_{19}$ system, respectively. Energy units are eV.}
	\centering
	\begin{tabular}{c c c c c c c c c}
		\hline
		System & \multicolumn{2}{c}{TD-DFT} & \multicolumn{2}{c}{TD-DFTB} & \multicolumn{2}{c}{TFW} & \multicolumn{2}{c}{LMGP+JP} \\
		          & pos  & shift & pos  & shift & pos  & shift & pos  & shift \\\hline
		Ag$_{19}$ & 3.56 &       & 2.67 &       & 2.50 &       & 2.66 &       \\
		Ag$_{25}$ & 3.27 & -0.29 & 2.58 & -0.09 & 2.39 & -0.11 & 2.51 & -0.15 \\
		Ag$_{31}$ & 3.01 & -0.55 & 2.41 & -0.26 & 2.26 & -0.24 & 2.36 & -0.30 \\
		Ag$_{37}$ & 2.77 & -0.79 & 2.26 & -0.41 & 2.15 & -0.35 & 2.22 & -0.44 \\
		Ag$_{43}$ & 2.57 & -0.99 & 2.08 & -0.59 & 2.05 & -0.45 & 2.11 & -0.55 \\
		Ag$_{49}$ & 2.41 & -1.15 & 2.02 & -0.65 & 1.95 & -0.55 & 2.01 & -0.65 \\
		Ag$_{55}$ & 2.30 & -1.26 & 1.91 & -0.76 & 1.85 & -0.65 & 1.91 & -0.75 \\
		Ag$_{61}$ & 2.04 & -1.52 & 1.74 & -0.93 & 1.71 & -0.79 & 1.76 & -0.90 \\\hline
	\end{tabular}
	\label{tab:Ag_rod}
\end{table}

\subsection{Benchmarking semiconductor systems}

\begin{figure}[htp]
	\centering
	\includegraphics[width=\textwidth]{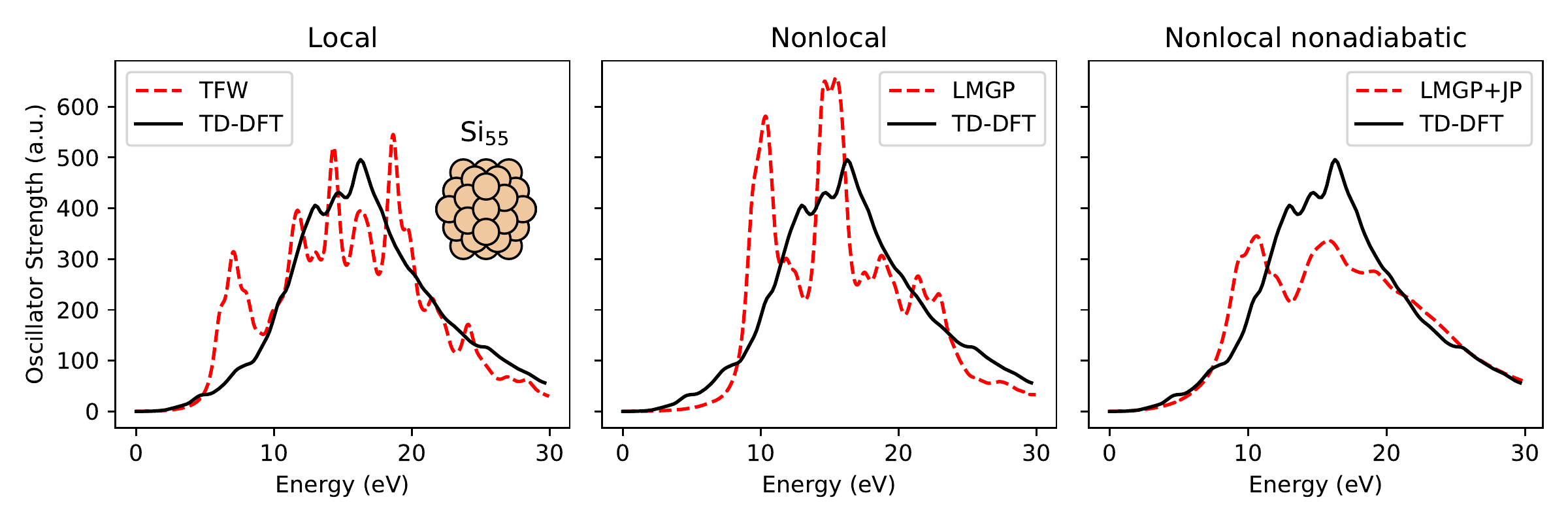}
	%\caption{Comparison of the effects of the nonlocal and nonadiabatic Pauli potential in the oscillator strength for icosahedron Si$_{55}$ cluster. The insect shows the structure of the cluster. Black solid: TD-DFT benchmark. Red dash-dotted: TFW (local). Green dashed: LMGP (nonlocal). Blue solid: LMGP+JP (nonlocal and nonadiabatic).}
	\caption{Optical spectrum of the Si$_{55}$ cluster. See caption to Figure \ref{fig:Mg8} for additional details.}
	\label{fig:Si55}
\end{figure}

We start the investigation of the optical spectra of clusters of semiconducting systems with a Si$_{55}$ icosahedral cluster carved in a way analogous to the metallic clusters mentioned in the previous section. Figure \ref{fig:Si55} compares the optical spectra of the Si$_{55}$ cluster. TFW and LMGP only qualitatively capture the spectra, with several peaks with overestimated oscillator strength. For LMGP, the peaks at $\sim10$ eV and $\sim15$ eV are too strong. The nonadiabatic JP correction improves the spectrum by lowering the oscillator strengths of these two peaks in favor of an increased intensity in the high-energy tail of the spectrum.

\begin{figure}[htp]
	\centering
	\includegraphics[width=0.70\textwidth]{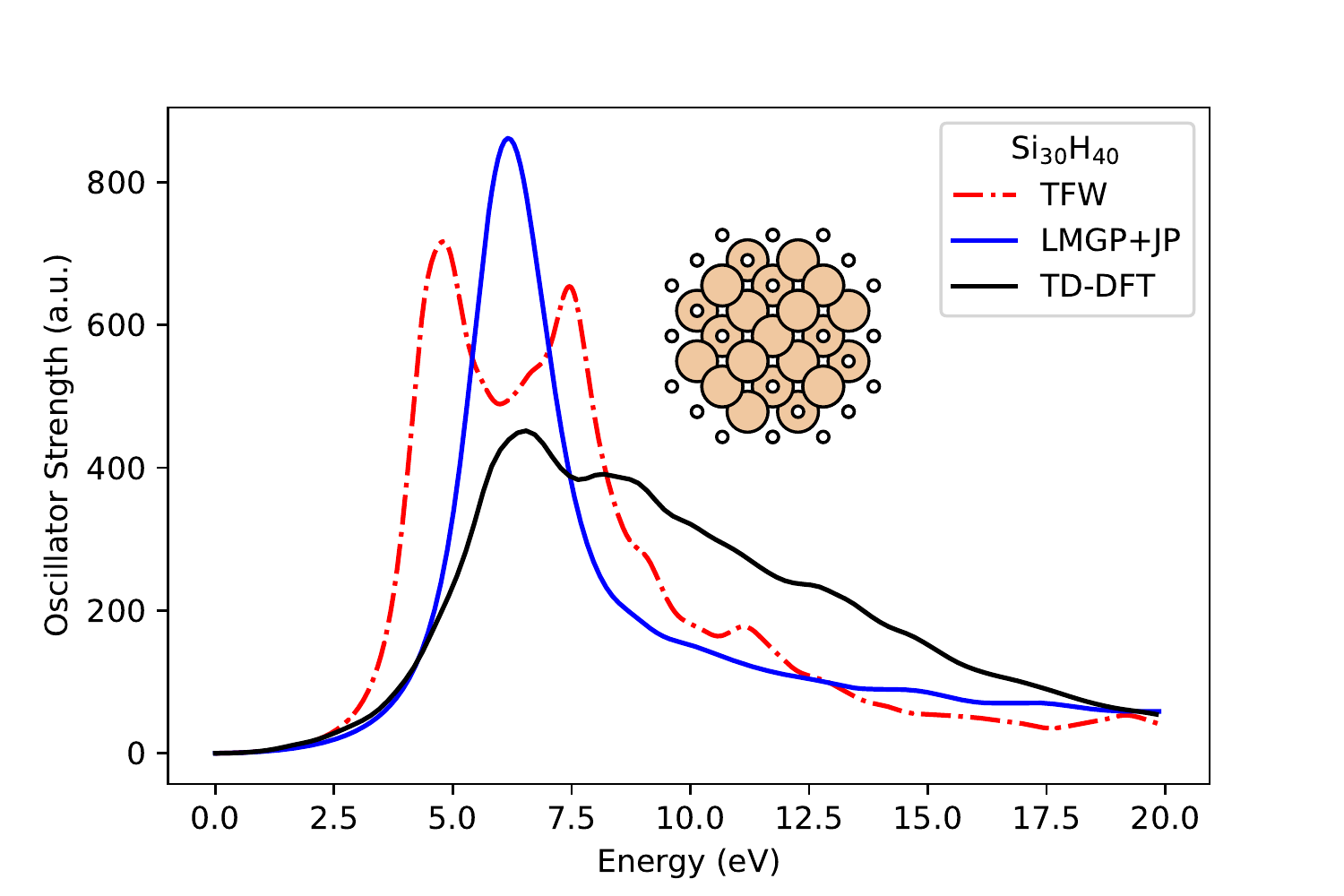}
	\caption{Optical spectrum of the Si$_{30}$H$_{40}$ cluster (see picture in the inset for the structure of the cluster). Solid black: TD-DFT benchmark. Dash-dotted red: TFW (local). Solid blue: LMGP+JP (nonlocal and nonadiabatic).}
	\label{fig:Si30H40}
\end{figure}

\begin{figure}[htp]
	\centering
	\includegraphics[width=0.875\textwidth]{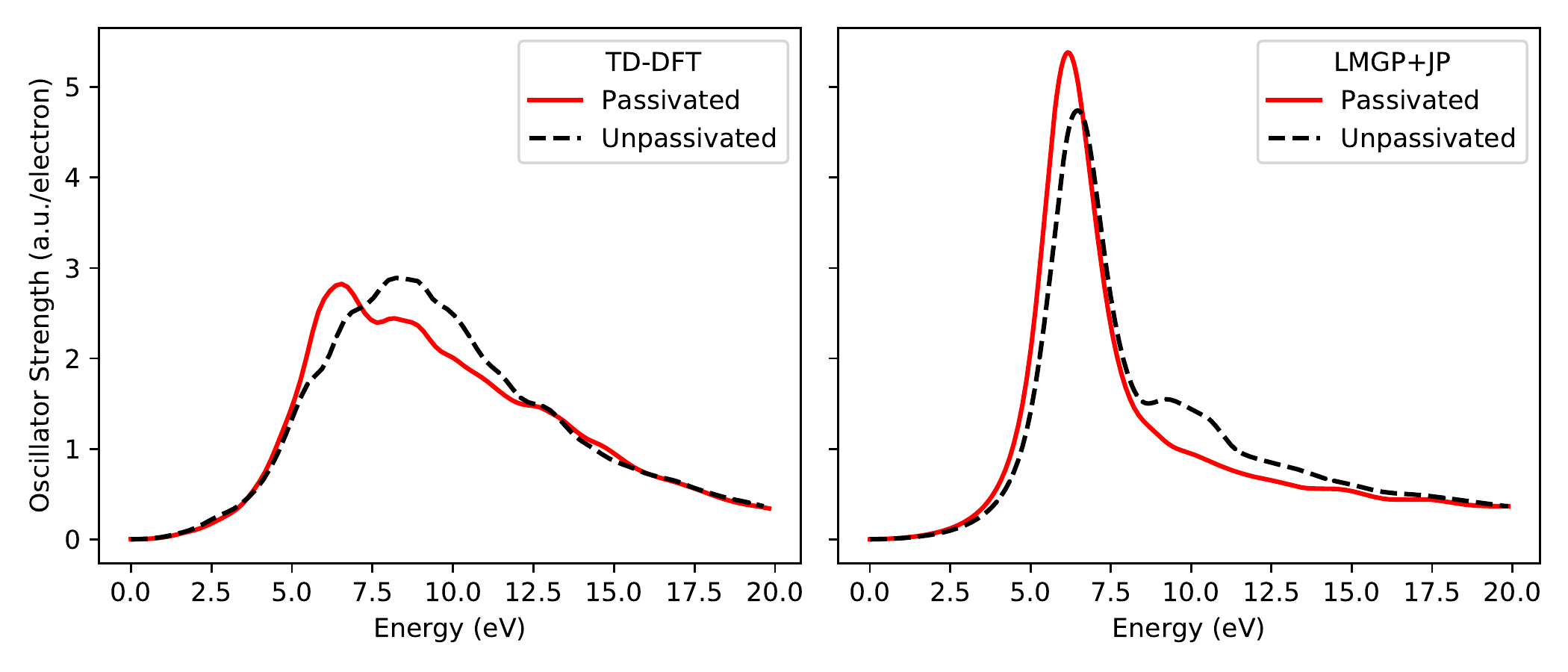}
	\caption{Effects of passivation with H atoms on the optical spectrum of Si$_{30}$. Left panel: TD-DFT. Right panel: TD-OFDFT carried out with with adiabatic LMGP plus nonadiabatic correction (LMGP+JP).}
	\label{fig:Sipass}
\end{figure}

We move on to consider an additional Silicon system, Si$_{30}$ cluster cut from a crystal diamond Si bulk structure. We also consider the H-passivated version of the cluster employing the passivation protocol described elsewhere \cite{HLC05} (i.e. for each dangling Si bond we put a hydrogen atom on the other end of the bond). We carefully chose the size of the structure so that each Si atom can have at most two Si-H bonds. In contrast to the icosahedron clusters which are more metallic, this configuration of the Si cluster is more semiconductor like.  The optical spectrum for this system is available in Figure \ref{fig:Si30H40}. TFW does not yield the correct peak position in the spectrum. Instead, LMGP+JP obtains the correct position of the lowest-lying peak while it blends away the other features of the spectrum. % The error is mostly due to the large error in the ground-state density of OFDFT, which proposes challenges in finding accurate excited states for TD-OFDFT. 

To interrogate TD-OFDFT's ability to predict trends when chemical changes are applied to the system, in Figure \ref{fig:Sipass} we compare the effect of passivation on the optical spectrum. TD-DFT predicts that the passivation induces a slight  red shift of the spectrum. Interestingly, a similar red shift can also be observed in the LMGP+JP TD-OFDFT result.
%We can see TD-OFDFT correctly captures the fact that passivation slightly red-shifts of the spectra. For the oscillator strength of the passivated system, the shapes of the spectra are a bit different, where TD-OFDFT has narrower and sharper peaks comparing to TD-DFT. However, TD-OFDFT correctly captures the two peaks feature with the first peak at $\sim$5eV for OF and $\sim$6eV for KS, and the second peak at $\sim$8eV for both methods.

%\subsection{GaAs}

\begin{figure}[htp]
	\centering
	\includegraphics[width=\textwidth]{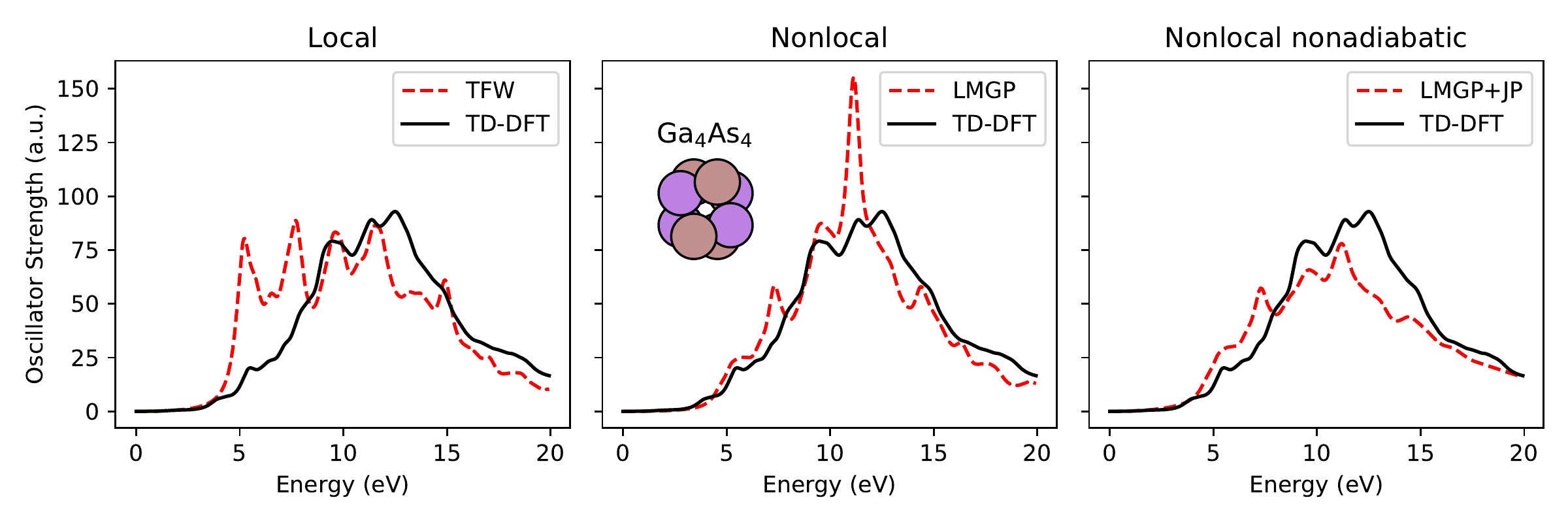}
	\caption{Optical spectrum of the Ga$_{4}$As$_{4}$ cluster. See caption to Figure \ref{fig:Mg8} for additional details.}
	%\caption{Comparison of the effects of the nonlocal and nonadiabatic Pauli potential in the oscillator strength for a Ga$_4$As$_4$ cluster. The insect shows the structure of the cluster.  Black solid: TD-DFT benchmark. Red dash-dotted: TFW (local). Green dashed: LMGP (nonlocal). Blue solid: LMGP+JP (nonlocal and nonadiabatic).}
	\label{fig:Ga4As4}
\end{figure}

OFDFT can  perform well for III-V semiconductors \cite{HC10,MGP18,LKT18,MP19a}. Figure \ref{fig:Ga4As4} compares the optical spectrum computed with TD-OFDFT and TD-DFT for a Ga$_4$As$_4$ cluster. Once again, by accounting for nonlocality and nonadiabaticity in the Pauli potential, the TD-OFDFT spectra are very similar to the TD-DFT benchmark result.
%Similar to that of the icosahedron Si cluster, TD-OFDFT correctly features the position of all peaks but the peaks are more pronounced. To illustrate that TD-DFT do have the two peaks at $\sim$5 and $\sim$8eV, we compare the peaks in a different way in fig. \ref{fig:Ga4As4_mu2}, which compares $|\delta\mu(\omega)|^2$, where $\delta\mu(\omega)$ is the dipole moment in the frequency space. Here we can see both TD-OFDFT and TD-DFT have the same number of peaks, except that TD-OFDFT has much higher peaks at low energy region.

%\begin{figure}[htp]
%	\centering
%	\includegraphics[width=0.70\textwidth]{figure/Ga4As4_mu2.pdf}
%	\caption{Comparison of $|\delta\mu(\omega)|^2$ power spectrum between TD-OFDFT and TD-DFT for Ga$_4$As$_4$ cluster, where $\delta\mu(\omega)$ is the dipole moment in the frequency space.}
%	\label{fig:Ga4As4_mu2}
%	
%\end{figure}
%\begin{figure}[htp]
%	\centering
%	\includegraphics[width=0.875\textwidth]{figure/GaAs_passivation.pdf}
%	\caption{Comparison of the effects of passivation in oscillator strength in a Ga$_{152}$As$_{152}$ cluster. Left panel: TD-OFDFT with TFW. Right panel: TD-OFDFT with LMGP+JP.}
%	\label{fig:GaAspass}
%\end{figure}

\subsection{Going beyond the capabilities of TD-DFT}

The optical spectrum of large nanoparticles is an important experimental characterization for optoelectronic materials \cite{W97}. We consider a Ga$_{152}$As$_{152}$ cluster, cut from the bulk GaAs structure, and passivated in a similar way as described previously for Si. Following a well-established recipe \cite{HLC05}, for GaAs quantum dots instead of passivating with H atoms, for each Ga atom a pseudo H atom with 1.25 electrons and for each As atom a pseudo H atom with 0.75 electrons are used. 

\begin{figure}[htp]
	\centering
	\includegraphics[width=0.875\textwidth]{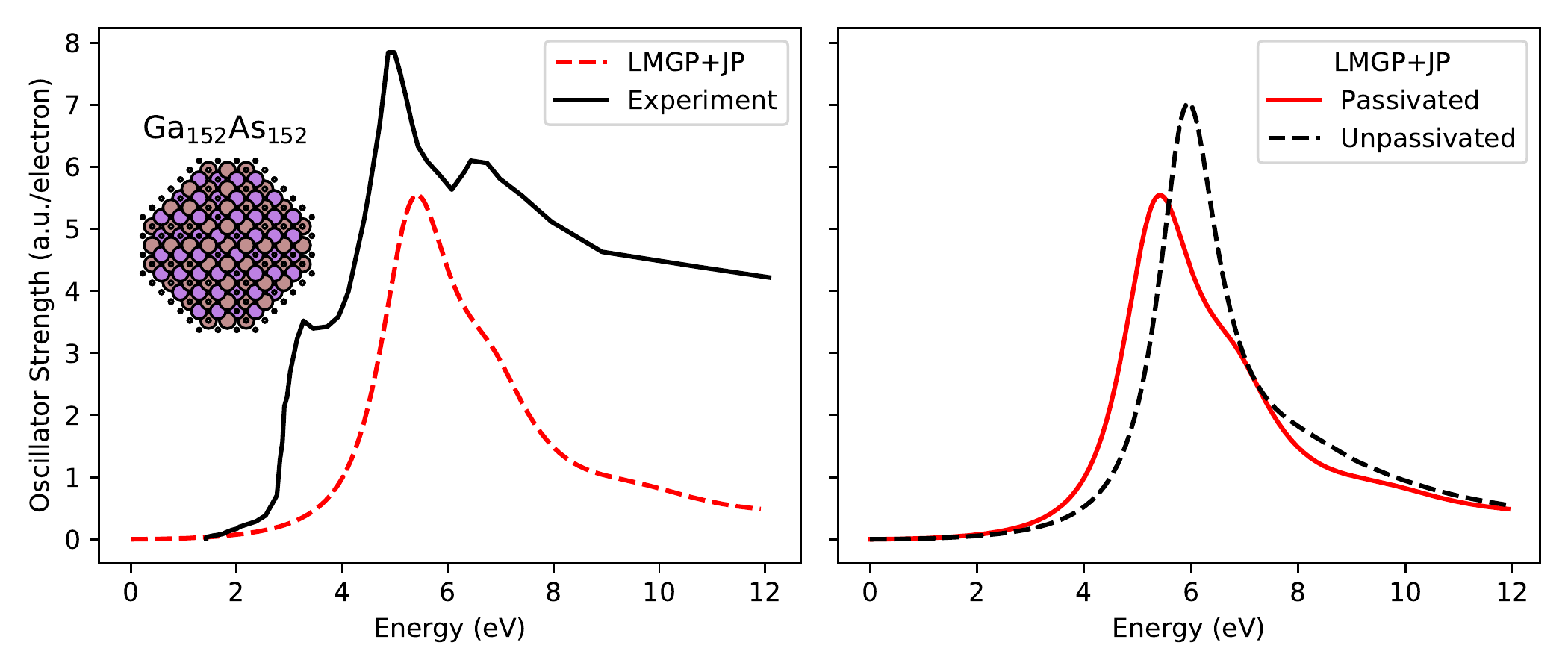}
	\caption{Left: TD-OFDFT optical spectrum of a passivated Ga$_{152}$As$_{152}$ cluster carried out with adiabatic LMGP plus nonadiabatic correction (LMGP+JP). Experimental spectrum is for bulk GaAs \cite{PE63}. Right: Effects of passivation on the optical spectrum of the cluster with TD-OFDFT.}
	\label{fig:Ga152As152}
\end{figure}

Figure \ref{fig:Ga152As152} illustrates the optical spectrum for the passivated and unpassivated cluster computed with the nonlocal and nonadiabatic method LMGP+JP in comparison with an experimental spectrum recorded for bulk GaAs. Tackling this system with TD-DFT would be a major undertaking wich would probably require specialized software and computing hardware. Our TD-OFDFT method was run on a single cluster node (36 cores). 

Interestingly, while TD-OFDFT recovers the main spectral feature at $\sim$ 5 eV as well as the smaller feature at $\sim$ 6 eV, it misses the band gap transition at $\sim$ 2.5 eV. Because OFDFT does not have a notion of single-particle orbitals or bands, it is not surprising that several details in the optical spectra related to interband trasitions are missing in TD-OFDFT. Formally, these would be recovered by the nonadiabatic Pauli potential. However, in this work, we have developed a nonadiabatic potential which was derived from the uniform electron gas. Thus, it would simply be too much to ask from the uniform electron gas to also be able to reproduce interband transitions in GaAs. Overall, despite the mentioned flaws, the agreement between the TD-OFDFT and the experiment is better than semiquantitative.

\subsection{Comparison against a truncated nonadiabatic Pauli potential}
\begin{figure}[htp]
	\centering
	\includegraphics[width=0.7\textwidth]{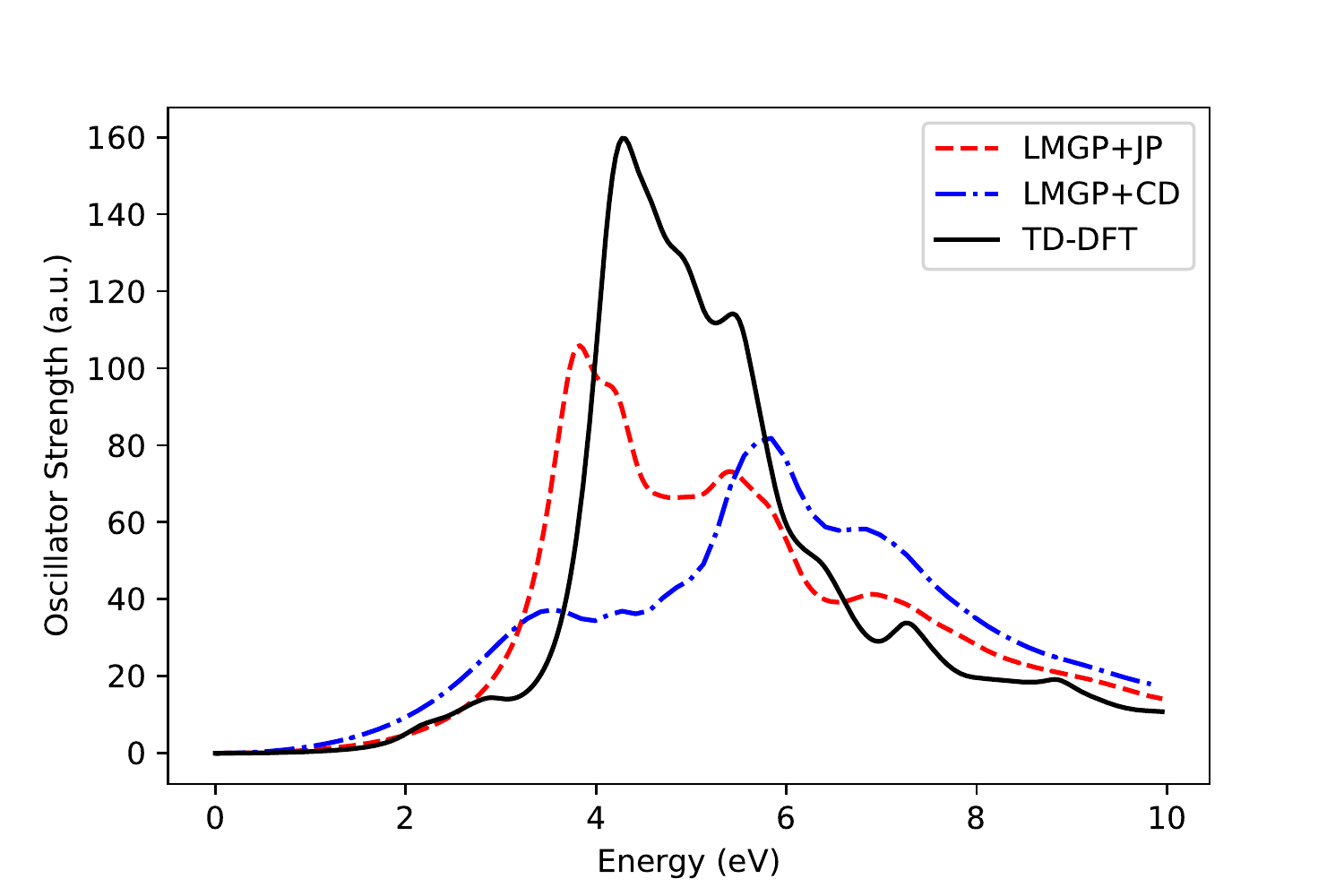}
	\caption{Optical spectrum of the Mg$_8$ cluster (same as in Figure \ref{fig:Mg8}) computed with TD-DFT and two versions of the nonadiabatic Pauli potential (see text for details).}
	\label{fig:Mg_cd}
\end{figure}

As mentioned, in Ref.\ \citenum{WCD+18} the authors pioneered the use a nonadiabatic Pauli potential  in real-time TD-OFDFT simulations. Their so-called ``current-dependent'' potential reads as follows (see Eq.\hspace{1pt}(15) of the reference),
\begin{equation}
	\label{eq:vp_cd}
	v\P^{\rm CD}(\br,t) = \frac{\pi^3}{2k_F^2(\br)}\cF^{-1}\left\{i\bq\cdot\bj(\bq,t)\frac{1}{q}\right\},
\end{equation}
which can be identified as the first term of \eqn{eq:vp_jp}. Here we wish to determine the effect of neglecting the second term on the optical spectra. The physical picture is that \eqn{eq:vp_cd} handles the low-$q$ dependence of the potential in the same way the Thomas-Fermi functional handles the low-$q$ density dependence of the kinetic energy functional. The second term of \eqn{eq:vp_jp}, instead, behaves in a way reminiscent of the von Weizs\"acker term (i.e., it goes like $q^2$). Therefore, in systems where large $q$ values are sampled (such as in clusters and any nonperiodic system) we expect the second term to play a significant role. 

In Figure \ref{fig:Mg_cd}, we report the optical spectra of Mg$_8$ computed with TD-DFT, and nonadiabatic TD-OFDFT with the two options for the nonadiabatic potential (LMGP+JP corresponds to \eqn{eq:vp_jp}, LMGP+CD corresponds to \eqn{eq:vp_cd}). From the figure, it is clear that the von Weizs\"acker-like term is instrumental in achieving semiquantitative agreement with the TD-DFT reference calculation. Once again, we believe the reason stems from the recovered $q^2$ behavior of the additional term present in \eqn{eq:vp_jp}.

\section{Conclusion}
We proposed and implemented a nonadiabatic (current-dependent) time-dependent Pauli potential for time-dependent orbital-free DFT (TD-OFDFT) simulations of electron dynamics in materials. The nonadiabatic part of the Pauli potential is responsible for major qualitative features in the optical spectrum because it bridges the dynamics of $N$ noninteracting bosons with the one of $N$ noninteracting electrons. The proposed potential derives from the frequency dependent dielectric function of the uniform electron gas. %Thus, it is expected to describe well the electron dynamics of metallic systems. For semiconductors and other systems where strong interband transitions take place, the proposed potential may only deliver semiquantitative spectra.

We test our method on clusters of metallic elements (Na, Mg, Ag) and semiconductors (Si, GaAs). In most cases, we find our TD-OFDFT to be close to bechmark TD-DFT spectra. TD-DFT is chosen as benchmark because it relies on an exact description of the $N$-electron fermionic system (i.e., exact time-dependent Pauli potential). The study of Na clusters revealed that previous investigations bare of the nonadiabatic local Pauli potential enjoyed strong error cancellation between the absence of nonadiabaticity in the potential and errors in the dynamics from the adiabatic part of the potential (which we address employing the LMGP nonlocal kinetic energy functional). As expected, the method struggles in the quantitative description of features in the spectra attributed to interband excitations. However, even for such difficult cases as GaAs nanoparticles, the method captures semiquantitatively the optical spectra and overall spectral envelopes. 

This work sets the stage for using time-dependent orbital-free methods for simulating the dynamics of complex systems for chemistry, energy applications and beyond.

\begin{acknowledgments} 
		This work is supported by the U.S.\ Department of Energy, Office of Basic Energy Sciences, under Award Number DE-SC0018343. The authors acknowledge the Office of Advanced Research Computing (OARC) at Rutgers, The State University of New Jersey for providing access to the Amarel and Caliburn clusters and associated research computing resources that have contributed to the results reported here. URL: http://oarc.rutgers.edu
\end{acknowledgments}
\bibliography{td-of-dft}
\end{document}